\documentclass[12pt]{article}
\usepackage{a4wide}
\usepackage{amssymb}
\begin{document}
{\renewcommand{\thefootnote}{\fnsymbol{footnote}}
%\hfill  IGC--yy/m--n\\
\medskip
\begin{center}
{\LARGE  Non-linear (loop) quantum cosmology}\\
\vspace{1.5em}
Martin Bojowald,$^1$ %\footnote{e-mail address: {\tt bojowald@gravity.psu.edu}},
Alexander L.~Chinchilli,$^1$
Christine C.~Dantas,$^2$\\
Matthew Jaffe$^3$
and David Simpson$^1$\\
% and ... 
%\\
\vspace{0.5em}
$^1$ Institute for Gravitation and the Cosmos,\\
The Pennsylvania State
University,\\
104 Davey Lab, University Park, PA 16802, USA\\
\vspace{0.5em}
$^2$ Divis\~ao de Materiais (AMR), Instituto de Aeron\'autica e Espaco (IAE),\\
Departamento de Ci\^encia e Tecnologia Aeroespacial (DCTA), Brazil\\
\vspace{0.5em}
$^3$ University of California, Berkeley, Department of Physics,\\
Berkeley, CA 94720, USA
\vspace{1.5em}
\end{center}
}

\setcounter{footnote}{0}

\newcommand{\lP}{\ell_{\mathrm P}}

\newcommand{\md}{{\mathrm{d}}}
\newcommand{\tr}{\mathop{\mathrm{tr}}}
\newcommand{\sgn}{\mathop{\mathrm{sgn}}}

\newcommand*{\R}{{\mathbb R}}
\newcommand*{\N}{{\mathbb N}}
\newcommand*{\Z}{{\mathbb Z}}
\newcommand*{\Q}{{\mathbb Q}}
\newcommand*{\C}{{\mathbb C}}

\begin{abstract}
  Inhomogeneous quantum cosmology is modeled as a dynamical system of discrete
  patches, whose interacting many-body equations can be mapped to a non-linear
  mini\-superspace equation by methods analogous to Bose--Einstein
  condensation. Complicated gravitational dynamics can therefore be described
  by more-manageable equations for finitely many degrees of freedom, for which
  powerful solution procedures are available, including effective
  equations. The specific form of non-linear and non-local equations suggests
  new questions for mathematical and computational investigations, and general
  properties of non-linear wave equations lead to several new options for
  physical effects and tests of the consistency of loop quantum gravity. In
  particular, our quantum cosmological methods show how sizeable quantum
  corrections in a low-curvature universe can arise from tiny local
  contributions adding up coherently in large regions.
\end{abstract}

\section{Introduction}

One of the main problems in deriving a reliable Planck-regime scenario in
canonical quantum cosmology is the question of how to include
inhomogeneity. While homogeneous models can easily be quantized, inhomogeneous
degrees of freedom severely complicate mathematical evaluations. Even the
formulation of consistent evolution equations, subject to the anomaly problem,
remains incomplete: no consistent and covariant version of inhomogeneous modes
valid at high density and including all relevant quantum effects is available
at present.

As possible solutions, two approaches have been developed so far, mainly with
the methods of loop quantum cosmology \cite{LivRev,Springer}. First, effective
equations have been successful in addressing the anomaly problem
\cite{ConstraintAlgebra,ScalarGaugeInv} and in including all relevant quantum
effects in sufficiently general form. (For details, see \cite{ReviewEff}.)
Potentially observable phenomena have been uncovered, showing physical
consequences of discrete quantum geometry
\cite{ScalarGaugeInv,SuperInflTensor,CHRev,BounceCMB,TensorHalfII,TensorSlowRollInv,TensorSlowRoll,GravWaveLQC,GCPheno}
and making the theory falsifiable \cite{InflConsist,InflTest}. At high
density, the implications of quantum space-time can be dramatic: general
properties of effective constrained systems show the presence of signature
change, turning Lorentzian space-time into a quantum version of 4-dimensional
Euclidean space \cite{JR,ScalarHol,Action,SigChange}. Bounces as they had
often been envisaged as non-singular versions of cosmology
\cite{BounceReview}, and formulated in quite some detail \cite{APS} in
homogeneous models of loop quantum cosmology, are then replaced by acausal
pieces of 4-dimensional space devoid of deterministic evolution. Regarding
specific field equations and details of the transition, however, present
calculations remain incomplete because not all quantum effects could yet be
implemented consistently at high density. Moreover, although state properties
can be derived by canonical effective equations, finding full quantum states
is difficult in this setting.

One of the alternatives is so-called hybrid quantization \cite{Hybrid}, in
which one combines a loop-quantized homogeneous background model with Fock
quantized inhomogeneous modes. Wave functions can then be solved for and
evolved, at least numerically and with certain truncations
\cite{InhomThroughBounce}. However, using a Fock quantization for
inhomogeneous modes, one does not directly deal with the discreteness of
space-time. Moreover, the hybrid method does not address the anomaly problem;
like related apporaches \cite{BKdustI,BKdustII,DeparamQG,ScalarRef}, it rather
avoids dealing with the problem by fixing the gauge or using
deparameterization, choosing a time variable before quantization. With these
additional steps, it is unlikely that the correct space-time picture is
obtained in covariant form, and in fact the models evaluated so far have
missed the signature change at high density.

In this article, we introduce a new way of incorporating effects of
inhomogeneity in loop quantum cosmology, dealing directly with wave
functions. Adapting ideas of condensed-matter physics used to describe
Bose--Einstein condensates,\footnote{Bose--Einstein condensates have been used
  in cosmological models before \cite{ChiralCondensate}, but for matter rather
  than quantum geometry. Moreover, we are not suggesting that there is
  physical condensation, but rather use related mathematical techniques to
  describe approximately homogeneous geometries.} some effects of
inhomogeneity will not be described by individual degrees of freedom but
rather by non-linearity of wave equations for homogeneous models. The
relationship between difference equations of loop quantum cosmology
\cite{cosmoIV,IsoCosmo} and certain integrable non-linear Schr\"odinger
equations has been noted in \cite{LQCSpin}, providing additional
motivation. The aim of the present article is to lay down the main ideas and
to point out several new consequences for quantum cosmology. We find that tiny
quantum corrections from inhomogeneous contributions to a large universe can
add up coherently to produce sizeable effects on average, to be included in
minisuperspace models. At this level, we will not yet address the anomaly
problem and derive a detailed model, but we will demonstrate the prospects for
this to be done at a later stage.

\section{Product states}

We start with a common way of dealing with inhomogeneity, viewing quantum
space at a given time (a spatial slice used in canonical quantum gravity) as a
collection of small homogeneous parts. As one moves between spatial slices,
the geometry evolves, resembling a many-body system of ``interacting''
elementary building blocks.  Each building block (called a patch) has a
quantum geometry described by a wave function of one of the well-known
homogeneous models of quantum cosmology, and they all interact dynamically
according to the quantized gravitational Hamiltonian.

\subsection{Classical model}

For simplicity, in this article we will assume isotropic patch geometries,
determined classically by a volume degree of freedom $V_{i,j,k}$ per patch,
labelled by integers $i$, $j$, $k$ to count patches in each spatial
direction. A given spatial slice $\Sigma= \bigcup_{i,j,k=1}^{{\cal N}^{1/3}}
{\cal V}_{i,j,k}$ --- a differentiable manifold with a local atlas of
coordinates --- is then the union of ${\cal N}$ patches ${\cal V}_{i,j,k}$, or
${\cal N}^{1/3}$ in each spatial direction. (Had we used anisotropic but still
homogeneous patches, we would in general have three independent factors in
${\cal N}={\cal N}_1{\cal N}_2{\cal N}_3$.) For now, we will assume ${\cal N}$
to be constant, which should be good for sufficiently brief evolution
times. In more realistic models, the number ${\cal N}$ of patches should
change in time, either by a fundamental process of discrete geometries being
refined \cite{InhomLattice,CosConst}, or by an approximation procedure akin to
adaptive mesh refinement that maintains the decomposition into isotropic
patches as a good model. (A time-dependent number of degrees of freedom is a
general problem, studied for instance in
\cite{Weiss,UnruhTime,River,EvolvingHilbert,CanSimp}.)

For simplicity, we choose coordinates in space such that each patch has the
same coordinate volume $\int_{{\cal V}_{i,j,k}} {\rm d}^3x = \ell_0^3$, with
$\ell_0^3=V_0/{\cal N}$ in terms of the total coordinate volume $V_0$ of
$\Sigma$ (or of a large compact subset). The geometrical volume of each patch
is then determined by the spatial metric which, if it is inhomogeneous, gives
rise to different patch volumes $V_{i,j,k}$. We assume that the metric is
close to the one of a spatially flat, isotropic model with a longitudinal
scalar mode, $h_{ab}= a(t)^2 \delta_{ab}+ 2L(t,x,y,z) \delta_{ab}$. (We will
use a lapse function corresponding to proper time, $N=1-2L/a^2$.) The patch
volumes then take the values
\begin{eqnarray}
 V_{i,j,k} &=& \int_{{\cal V}_{i,j,k}} {\rm d}^3x \sqrt{\det h} = a^3
 \int_{{\cal V}_{i,j,k}} {\rm d}^3x (1+2L/a^2)^{3/2}\nonumber \\
 &\approx& a^3\ell_0^3+
 3a\int_{{\cal V}_{i,j,k}} {\rm d}^3x L \approx \frac{V}{{\cal N}}
 +3aL(x_{i,j,k})\ell_0^3 \label{VL}
\end{eqnarray}
with the total volume $V=a^3V_0=a^3\ell_0^3{\cal N}$.  In the two
approximations in the second line of this equation, we have first expanded the
root and then replaced the patch-integrated $L$ by its value at a point
$x_{i,j,k}\in {\cal V}_{i,j,k}$, such as the center. Since we assume the
patches to be nearly isotropic and smaller than the variation scale of the
perturbative inhomogeneity, $L$, both approximations are well
justified. Solving (\ref{VL}) for $L$, we can therefore replace the continuum
function $L$ by deviations of the discrete variables $V_{i,j,k}$ from the
total volume $V={\cal N}a^3\ell_0^3$:
\begin{equation}
 L(x_{i,j,k}) = \frac{V_{i,j,k}-V/{\cal N}}{3a\ell_0^3}\,.
\end{equation}

The dynamics of the $V_{i,j,k}$ as functions of time is governed by a
discretized version of the Hamiltonian constraint 
\begin{equation}
 H_{\rm grav}+H_{\rm matter}=0
\end{equation} 
of general relativity, with contributions from the gravitational field and
from matter. At this point, one will eventually have to face the problem of
time and the anomaly problem.\footnote{Regimes of interest here, with small
  inhomogeneity, are usually semiclassical regarding quantum geometry. In such
  a situation, effective constraints \cite{EffCons,EffConsRel,EffConsComp} can
  be used to solve the problems of time
  \cite{EffTime,EffTimeLong,EffTimeCosmo} and anomalies
  \cite{ConstraintAlgebra}.} In this article, however, we focus on laying out
the details of the new model, and therefore circumvent these difficult
problems by formulating the dynamics in a specific gauge. With this choice, we
may be blind to the complete quantum space-time structure, but new qualitative
effects should still become visible. To proceed and to be specific, we assume
matter to be dust, with Hamiltonian $H_{\rm matter}= p_t/a^3$, where $p_t$ is
a momentum variable conjugate to a matter degree of freedom $t$ that will play
the role of time. The role of time is made clear if we rewrite the Hamiltonian
constraint equation as
\begin{equation}
 p_t = -a^3 H_{\rm grav}= -\frac{V}{\ell_0^3} H_{\rm grav}\,.
\end{equation}
The variable $p_t$ then appears formally as an energy, or a canonical
Hamiltonian that generates evolution with respect to $t$. (More generally, we
could assume matter to contribute to the Hamiltonian constraint by $H_{\rm
  matter}= p_{t'}/a^{3(1+w)}$ if there is a perfect fluid with
equation-of-state parameter $w$. A time variable $t'$ different from $t$ then
parameterizes evolution.)

To derive the dynamics in detail, we start with the classical Hamiltonian
constraint of general relativity and write it in discrete canonical variables
$V_{i,j,k}$ together with their momenta $\Pi_{i,j,k}$, related to
$\dot{V}_{i,j,k}$. In the ADM formulation of canonical gravity, the spatial
metric $h_{ab}$ is canonically conjugate to
\begin{equation} \label{piab}
 \pi^{ab}= \frac{\sqrt{\det h}}{16\pi G} (K^{ab}- K^c_ch^{ab})\,.
\end{equation}
(See \cite{CUP} for an introduction to canonical gravity.)  We compute the
canonical variables in our perturbed situation by writing $h_{ab}=
\bar{h}\delta_{ab}+\delta h_{ab}$ and $\pi^{ab}=
(\bar{\pi}/V_0)\delta^{ab}+\delta \pi^{ab}$, split into background variables
$\bar{h}=a^2$ and $\bar{\pi}$ (spatial constants) and inhomogeneity $\delta
h_{ab}$ and $\delta\pi^{ab}$. We divide $\bar{\pi}$ by $V_0$ in $\pi^{ab}$ to
ensure that the symplectic term $\int_{\Sigma}{\rm d}^3x \dot{h}_{ab}\pi^{ab}=
\dot{\bar{h}}\bar{\pi}+\cdots$ assumes the canonical form in its background
term.  For scalar modes in longitudinal gauge, $\delta h_{ab}=2L\delta_{ab}$
and $\delta\pi^{ab}= \delta\pi \delta^{ab}$.

To avoid overcounting of degrees of freedom, we require the inhomogeneity
$\delta f$ of any field $f=\bar{f}+\delta f$ to satisfy $\int_{\Sigma}{\rm
  d}^3x \delta f=0$ when integrated over all of space. (We turn
inhomogeneities of tensor fields such as $\delta h_{ab}$ into scalars using
the background metric $\delta_{ab}$.) As a consequence,
$\bar{f}=\int_{\Sigma}{\rm d}^3x f$ is indeed the spatial average.  At this
stage we do not assume that $\delta f$ is of first or any specific order in
perturbation theory; we have simply rearranged our degrees of freedom by
splitting them into background variables and inhomogeneity. The symplectic
structure, our current interst, only refers to degrees of freedom but not to
orders of perturbation theory: higher perturbative orders do not introduce new
degrees of freedom. We will introduce the perturbative expansion when we
prepare our Hamiltonian for a derivation and analysis of equations of motion.

Any terms linear in $\delta h_{ab}$ or $\delta\pi^{ab}$ in the Hamiltonian or
symplectic term $\int_{\Sigma} {\rm d}^3x \dot{h}_{ab}\pi^{ab}$ therefore
vanish. Inserting the inhomogeneous metric $h_{ab}=(a^2+2L)\delta_{ab}$ in
(\ref{piab}) (with vanishing shift and longitudinal lapse in $K_{ab}$) results
in the inhomogeneous momentum
\begin{equation}
 \pi^{ab}= -\frac{1}{8\pi G} \left(\dot{a}+
   a\left(\frac{L}{a^2}\right)^{\bullet}\right) \delta^{ab}
\end{equation}
from which we read off the momentum of $\bar{h}=a^2$ as
$\bar{\pi}=-\dot{a}V_0/8\pi G$ with the total coordinate volume $V_0$, and the
momentum of $\delta h_{ab}=2L\delta_{ab}$ as $\delta\pi^{ab}=
-(a(L/a^2)^{\bullet}/8\pi G)\delta^{ab}$. By a canonical transformation we can
switch to volume variables as defined in our patch model: we have momenta
\begin{equation} \label{Momenta}
 \Pi_V= -\frac{1}{12\pi G}\frac{\dot{V}}{V} \quad\mbox{ and }\quad
 \Pi_{i,j,k} = -\frac{1}{12\pi G} \left(\frac{{\cal
   N}V_{i,j,k}}{V}\right)^{\bullet}
\end{equation}
of $V$ and $V_{i,j,k}$.

For small inhomogeneity, it is sufficient to expand the Hamiltonian constraint
to second order in $\delta h_{ab}$ (or $L$) and its time and space
derivatives. Starting from
\begin{equation}
H_{\rm grav}=\frac{1}{16\pi G} 
\int_{\Sigma} {\rm d}^3xN\left(K^{ab}
  K_{ab}-K^{2}-{}^{3}\!R\right)\sqrt{\det h}=
\int_{\Sigma} {\rm d}^3x\mathcal{H}_{\rm grav}\,,
\end{equation}
we obtain
\[
\mathcal{H}_{\rm grav} \approx -\frac{3}{8\pi G}
\left(a\dot{a}^{2}+\frac{\dot{L}^{2}
 -4(\dot{a}/a)\dot{L}L
  +4(\dot{a}/a)^2L^{2}}{a}  +a^{-3} \sum_{b=1}^{3}\left(\left(\frac{\partial
        L}{\partial         x^{b}}\right)^{2} -\frac{4}{3} 
    L\left(\frac{\partial^{2} L}{\partial x^{b^{2}}}\right)\right)
            \right)\,.
\]
In the Hamiltonian $H_{\rm grav}=\int_{\Sigma}{\rm d}^3x \mathcal{H}_{\rm
  grav}$ we can integrate by parts in spatial derivatives, replacing
second-order derivatives by first-order ones. (Boundary terms will play no
role in what follows.) Moreover, it turns out that the time derivatives of $L$
can be written more compactly if we use $L/a^2$, a combination of variables
that is also more convenient when expressed by patch volumes: $ \dot{L}^{2}
-4(\dot{a}/a)\dot{L}L +4(\dot{a}/a)^2L^{2}= ((L/a^2)^{\bullet})^2$. The
Hamiltonian density we use will therefore be
\begin{equation}
\mathcal{H}_{\rm grav} = \frac{3a^3}{8\pi G}
\left(\left(\frac{\dot{a}}{a}\right)^{2}
+\left(\left(\frac{L}{a^2}\right)^{\bullet}\right)^2
 +\frac{7}{3}\frac{1}{a^2} \sum_{b=1}^{3}\left(\frac{\partial
        (L/a^2)}{\partial         x^{b}}\right)^{2} \right)  \,. \label{HExp}
\end{equation}

We then introduce our background momentum
$\dot{a}/a=-4\pi G \Pi_V$ and the patch momenta
$(L/a^2)^{\bullet}\rightarrow -4\pi G\Pi_{i,j,k}$ after replacing the
integral by a sum over patches, $\int_{\Sigma}{\rm d}^3x \mathcal{H}_{\rm
  grav}\approx \sum_{i,j,k} \mathcal{H}_{i,j,k}$. Our discretized Hamiltonian
then is
\begin{equation}
  H_{\rm grav}^{\rm disc}= -6\pi G V\left(\Pi_V^2 + \frac{1}{{\cal N}}
    \sum_{i,j,k}\Pi_{i,j,k}^2+ \cdots\right)
\end{equation}
where the dots indicate the derivative terms after discretization.

We have quadratic single-patch Hamiltonians in the first two terms, analogous
to harmonic 1-particle Hamiltonians of our many-body problem.  Spatial
derivatives of $L$ must be discretized before they can be expressed in terms
of the $V_{i,j,k}$. The discretization procedure is a matter of choice and, to
some degree, convenience; we will make use of
\begin{equation}
 \frac{\partial}{\partial x^b} \frac{L(x_{i,j,k})}{a^2} 
 \longrightarrow \frac{V_{(i,j,k)+\tilde{b}}-
   V_{(i,j,k)-\tilde{b}}}{6\ell_0 (V/{\cal N})}\,,
\end{equation}
indicating by $\tilde{b}$ the unit vector in the
$b$-direction.\footnote{Choices in the discretization procedure affect
  physical implications and can therefore be tested for their
  consistency. These choices are related to quantization and regularization
  ambiguities in canonical quantum gravity from which our expressions should
  follow in some complicated way. If discretization choices can be restricted,
  the same will be true for ambiguities of an underlying fundamental theory.}

Quadratic expressions of spatial derivatives in (\ref{HExp}) then provide
interaction terms that can be written as depending on either the patch
geometries in product form, such as
$V_{(i,j,k)+\tilde{b}}V_{(i,j,k)-\tilde{b}}$, or more conveniently, the
difference $(V_{(i,j,k)+\tilde{b}}- V_{(i,j,k)-\tilde{b}})$ in discrete
minisuperspace. The latter version is closer to interactions of many-body
systems depending on the distance between particles.

In addition to interactions between neighboring patches, each patch volume
interacts with the average volume $V$ because it appears in some factors in
the Hamiltonian. These variables are not independent but satisfy
$\sum_{i,j,k}V_{i,j,k}=V$. In order to focus on the self-interaction of
inhomogeneity, we will treat $V$ as an external parameter for the dynamics of
the $V_{i,j,k}$, corresponding to the common approximation in cosmology that
ignores back-reaction of inhomogeneity on the background.

\subsection{Quantization}

Each patch of volume $V_{i,j,k}$ and expansion rate related to $\Pi_{i,j,k}$
is isotropic and may be quantized as a single minisuperspace model,
corresponding to the 1-particle Hilbert space of a many-body system. One may
follow either Wheeler--DeWitt quantization or loop quantization, both with
volume representations in which $V_{i,j,k}$ becomes a multiplication
operator. In the former case, one deals with wave functions $\psi(V_{i,j,k})$
in $L^2({\mathbb R}_+,{\rm d}V_{i,j,k})$ and the momenta act by
$\hat{\Pi}_{i,j,k}= -i\hbar {\rm d}/{\rm d}V_{i,j,k}$; in the latter,
$\psi_{V_{i,j,k}}$ is an element of the non-separable sequence space
$\ell^2({\mathbb R})$ and exponentials of momenta, rather than momenta
themselves, are quantized:
\begin{equation} \label{shift}
\widehat{\exp(i\delta_{i,j,k}\Pi_{i,j,k}/\hbar)}\psi_{V_{i,j,k}}=
  \psi_{V_{i,j,k}+\delta_{i,j,k}} 
\end{equation}
for real numbers $\delta_{i,j,k}$ (whose values are to be fixed as part of
quantization choices).\footnote{We work in a special Abelian sector of
  homogeneous loop quantum cosmology. The general non-Abelian structure is
  more complicated due to refinement features \cite{NonAb}, but qualitative
  aspects are shown well by the Abelian simplification.}  The action of
$\widehat{\exp(i\delta_{i,j,k}\Pi_{i,j,k}/\hbar)}$ on the sequence space is
not continuous in $\delta_{i,j,k}$, and a derivative by $\delta_{i,j,k}$,
which would otherwise result in an operator for $\Pi_{i,j,k}$, does not
exist. (A second difference between the quantizations is that $V_{i,j,k}$ in
Wheeler--DeWitt models is usually taken as the (positive) volume, while loop
quantum cosmology is based on triad variables in which $V_{i,j,k}$ is the
oriented volume, which can turn negative if the orientation is reversed. We
therefore use the full real line ${\mathbb R}$ in the sequence space, rather
than ${\mathbb R}_+$. Note that this resolves self-adjointness issues of
derivative operators on $L^2({\mathbb R}_+,{\rm d}V_{i,j,k})$.)

Both representations are well-defined but not unitarily related to each other;
they lead to different physics. Especially at high curvature, where
$\Pi_{i,j,k}$ is large, effects of the loop quantization can differ
significantly from those of the Wheeler--DeWitt quantization. The discreteness
inherent in shift operators (\ref{shift}) relating derivatives is then
important, in addition to the discreteness implemented by our treatment of
inhomogeneity.

If inhomogeneity is small, the patches evolve nearly independently of one
another without strong correlations, and the evolved state remains a product
state $\Psi(V_1,V_2,\ldots)=\psi_1(V_1)\psi_2(V_2)\cdots$ of the individual
patch wave functions $\psi_i$ if the initial state is of such a form. Each
single-patch wave function evolves according to a differential
(Wheeler--DeWitt \cite{DeWitt}) or difference (loop quantum cosmology
\cite{cosmoIV,IsoCosmo}) equation if inhomogeneity can be ignored. With
inhomogeneity included, interaction terms between the individual wave
functions occur on superspace, complicating the dynamics. If inhomogeneity is
sufficiently small, however, the interactions can be treated by approximation,
such as perturbation theory.

Small inhomogeneity at the level of quantum geometry also implies that the
individual wave functions are very similar to one another, so that the full
state can approximately be written as
$\Psi(V_1,V_2,\ldots)=\psi(V_1)\psi(V_2)\cdots$ with a single wave function
$\psi$ to be solved for. This form of product states allows one to map
many-body dynamics to 1-particle dynamics in a specific potential, described
by a wave equation that turns out to be non-linear. At this stage, standard
techniques to describe matter condensates, in which individual wave functions
of different particles are exactly equal to one another, can be applied.

\subsection{Condensate}

By our preceding considerations in cosmology, we have realized a mathematical
formulation with all the ingredients used in the description of Bose--Einstein
condensation. We interrupt our discussion of cosmology to recall salient
features of this important system in condensed-matter physics. In this
example, $\Psi$ is a many-body state, and $\psi$ the 1-particle wave
function common to all constituents of the condensate. Taking the same $\psi$
is not an assumption because condensed particles have exactly the same wave
function.

Assuming pointlike interactions between the particles, described by a
delta-function potential of strength $\alpha$, we have the many-body
Hamiltonian
\begin{equation}
  \hat{H}= \sum_{i=1}^n \left(\frac{1}{2m} \hat{p}_i^2+ V(\hat{x}_i)\right)+
  \frac{1}{2}\alpha   \sum_{i\not= j} \delta(\hat{x}_i-\hat{x}_j)
\end{equation}
for $n$ particles of mass $m$ in individual potentials $V(x_i)$. With a
product state $\Psi(x_1,x_2,\ldots)=\psi(x_1)\psi(x_2)\cdots$ for the
condensate, we compute the expectation value of the Hamiltonian as
\begin{equation} \label{ExpH}
 \langle\hat{H}\rangle_{\Psi}= n \langle \hat{p}^2/2m+
 V(\hat{x})\rangle_{\psi}+ \frac{1}{2} n(n-1) \alpha \int {\rm d}^3x
 |\psi(x)|^4\,.
\end{equation}
The first term just adds up the 1-particle expectation values computed for the
wave function $\psi$. The second term is not equal to a 1-particle expectation
value. However, we can formally interpret it as the expectation value of a
``potential'' $|\psi(x)|^2$ depending on the wave function. Accordingly, the
1-particle dynamics and energy spectra are governed by a non-linear
Schr\"odinger equation, the Gross--Pitaevski equation
\begin{equation} \label{GP}
i\hbar\frac{\partial\psi}{\partial t}=
-\frac{\hbar^2}{2m}\frac{\partial^2\psi}{\partial x^2}+
 V(x)\psi + \frac{1}{2}(n-1)\alpha|\psi(x)|^2\psi\,.
\end{equation}
For a full and rigorous derivation, see \cite{KineticEqs,NonLinSchroed}.

Interacting many-body dynamics of the condensate wave function can therefore
be mapped to non-linear 1-particle dynamics. 

\section{Non-linear dynamics in quantum cosmology}

With the preparations presented in the preceding section, we propose a new
method to deal with small cosmological inhomogeneity, making use of the same
ideas and initial constructions employed to describe matter
condensates. Well-established methods then provide a tractable approximate
description by non-linear dynamics of a homogeneous model.

\subsection{Equation of motion}

Except for the differences in the conceptual nature, regarding for instance
the approximations and assumptions used, our model for inhomogeneous quantum
cosmology so far resembles those of matter condensates rather closely. The
main mathematical difference lies in the interaction potential. For particles
in a condensate, a delta function of the distance between particles is a good
approximation for nearly pointlike interactions, which can be smeared out to
more-complicated functions for realistic systems. The interaction potential we
obtain in cosmology, expanding and discretizing the gravitational Hamiltonian
constraint, is a quadratic polynomial in the distances in
minisuperspace. Although the single-patch wave equation we obtain is still
non-linear, as in the presence of any kind of interactions, it is more
complicated than in the Gross-Pitaevski equation.

Another difference between the models is the discreteness of the quantum
representation used in a loop quantization, in addition to the discretization
of space by patches ${\cal V}_{i,j,k}$. Not only space but also superspace is
then discrete. As a consequence, wave equations in loop quantum cosmology are
difference equations, and with our method to include inhomogeneity we will be
dealing with some version of a discrete non-linear Schr\"odinger equation,
one example given by
\begin{equation} \label{DiscNonLinS}
  i\hbar\frac{\partial\psi_n}{\partial t}= \frac{1}{2}(\psi_{n+1}-2|\psi_n|^2
\psi_n+  \psi_{n-1})\,.
\end{equation}
However, since we are not dealing with pointlike interactions in superspace,
modeled by delta functions, but rather with polynomials, the non-linearity
will be different. In fact, our equation will not only be non-linear but also
non-local but nevertheless, as it turns out, well-suited to canonical
effective methods.

Using the same starting point as in Bose--Einstein condensation, the key step
is to evaluate the expectation value of the interaction Hamiltonian in a
product state. To illustrate the main consequence, we consider just two
variables $V_1$ and $V_2$ interacting with each other via a potential $W_{\rm
  int}(V_1,V_2)= \alpha (V_1-V_2)^2/V^2$ as in a discretized (\ref{HExp}). We
divide by the total volume squared, treated as an external but
time-dependent parameter, in order to have the correct scaling behavior of the
Hamiltonian under a change of the spatial region. The expectation value of the
quantized $W_{\rm int}$ then produces a term
\begin{eqnarray}
 \langle\hat{W}_{\rm int}\rangle_{\Psi} &=&\frac{\alpha}{V^2} \int {\rm
   d}V_1{\rm d}V_2  |\psi(V_1)|^2 |\psi(V_2)|^2 (V_1-V_2)^2 \\
 &=& \frac{\alpha}{V^2} \int {\rm d}V_1 |\psi(V_1)|^2 \int {\rm d}\delta V
|\psi(V_1+\delta V)|^2 (\delta V)^2
\end{eqnarray}
where we introduce $\delta V:= V_2-V_1$. 

We can perform the second integration independently of the first over
$V_1$. It depends on the wave function, but if we assume that $\psi$ is
sharply peaked around the expectation value $\langle V_1\rangle$, the dominant
contribution to $\langle\hat{W}_{\rm int}\rangle_{\Psi}$ comes from values of
$V_1$ for which the second integration
\begin{equation}
 \int {\rm d}\delta V
|\psi(\langle V\rangle+\delta V)|^2 (\delta V)^2=(\Delta V)^2
\end{equation}
equals the quantum fluctuation of $V$ in the state $\psi(V)$. Instead of a
non-linearity potential depending on $\psi(V)$ or $\psi_n$ as in
(\ref{DiscNonLinS}), we have a non-linearity potential that depends on the
wave function via moments such as $\Delta V$. For instance, following the
preceding arguments and noting that the minisuperspace $V$ is quantized to a
discrete parameter $n$, we need to consider an equation of the form
\begin{equation} \label{NonLin}
 i\hbar \frac{\partial\psi_n}{\partial t} =
 \psi_{n+1}-2\left(1-\frac{1}{2}\alpha \frac{(\Delta n)_{\psi}^2}{n}\right)
 \psi_n +\psi_{n-1}\,.
\end{equation}
We note that equation (\ref{NonLin}) is not only non-linear but also
non-local: the coefficient $(\Delta n)_{\psi}^2=\sum_n (n-\langle
n\rangle_{\psi})^2|\psi_n|^2$ depends on all values of $\psi_n$. Moreover, the
equation as written is meaningful only for $n\not=0$. At $n=0$, the volume
vanishes and we encounter a cosmological singularity. By inverse-triad
corrections \cite{InvScale}, loop quantum cosmology resolves this singularity
in such a way that $1/n$ is replaced by a bounded function. For simplicity, we
will not discuss these terms here and instead focus on evolution at large $n$.

We must ensure that our assumption of a sharply peaked state remains true for
the approximation to be valid.  If the state is not sharply peaked or if the
approximation is to be driven to higher orders, we can use a derivative
expansion of $\psi$. Writing
\[
 |\psi(V_1+\delta V)|^2= |\psi(\langle V\rangle+\delta V+ (V_1-\langle
 V\rangle))|^2
\]
and expanding by $V_1-\langle V\rangle$, we obtain 
\[
 \langle\hat{W}_{\rm int}\rangle_{\Psi}= \int{\rm d}V_1 |\psi(V_1)|^2 W_{\rm
   nonlin}(V_1)
\]
with the non-linearity potential
\begin{equation} \label{WNonLin}
 W_{\rm nonlin}(V) = \sum_{j=0}^{\infty} \frac{1}{j!} (\Delta V)^2_{\rho^{(j)}}
 (V-\langle V\rangle)^j
\end{equation}
where the moment $(\Delta V)^2_{\rho^{(j)}}$ is the $V$-fluctuation computed
with the ``distribution'' $\rho^{(j)}$, defined as the $j$-th derivative of
$\rho(V)=|\psi(V)|^2$. Note that these derivatives need not be normalized or
positive, so that we do not have probability distributions and fluctuations in
the statistical sense. Nevertheless, the resulting numbers are well-defined as
parameterizations of the non-linearity potential.

We continue with a discussion of the leading-order equation (\ref{NonLin}).

\subsection{Solution procedures}

An inverse scattering transform is the method of choice to solve non-linear
discrete or differential Schr\"odinger equations \cite{NonLinDiff}. However,
the equation we obtain here, (\ref{NonLin}), is not only non-linear but also
non-local. Standard techniques are therefore not readily available.

Non-local equations can sometimes be treated by replacing the non-local
coefficient by new auxiliary degrees of freedom, as in \cite{NonLinLocal} in
the context of the non-linear Schr\"odinger equation. If the new degree of
freedom is subject to a differential or difference equation with a source term
given by the original wave function $\psi$, its general solution is a
non-local expression in $\psi$ (integrating its product with the Green's
function of the auxiliary equation). If the right equation is chosen, the
general solution for the auxiliary variable may provide the non-local
coefficient, $(\Delta n)^2$ in our case.  Here, however, such a treatment is
not obvious.

Instead, canonical effective methods \cite{EffAc} based on the dynamics of
moments of a state provide solution techniques well-suited for equations such
as (\ref{NonLin}). The non-local coefficient is a second-order moment of the
wave function; using equations for the moments instead of $\psi_n$ itself then
provides a reformulation of the problem in variables in which the non-locality
disappears. Morally, this procedure is a version of introducing new degrees of
freedom related to the wave function non-locally, for moments\footnote{Our
  notation is a variation of the common $(\Delta n)^2=\Delta(n^2)$ at second
  order, the parenthesis displaced to be unambiguous at higher orders.} such
as $\Delta(n^a):=\sum_n (n-\langle n\rangle)^a|\psi_n|^2$ with the expectation
value $\langle n \rangle=\sum_n n|\psi_n|^2$ are non-local in
$\psi_n$. However, in quantum physics the moments are not auxiliary variables
but rather variables of prime physical interest. For $a=2$, we have quantum
fluctuations, and higher moments with $a>2$ provide additional statistical
information about the state.

For linear discrete or differential Schr\"odinger equations, canonical effective
techniques \cite{EffAc} amount to a systematic expansion of Ehrenfest's
equations, used not just to derive the semiclassical limit in rigorous terms
\cite{Hepp} but also to compute quantum corrections to any desired order in
$\hbar$. For our purposes, we need to generalize these methods to non-linear
equations as encountered here.

In quantum mechanics, a set of $N$ basic operators $\hat{J}_i$ with closed
linear commutators 
\begin{equation} \label{JJ}
 [\hat{J}_i,\hat{J}_j]=\sum_k C_{ij}{}^kJ_k
\end{equation}
(perhaps including the identity operator if some commutators are constants)
provides a closed algebra for expectation values under Poisson brackets
\begin{equation}\label{Poisson}
 \{\langle\hat{J}_i\rangle,\langle\hat{J}_j\rangle\}=
 \frac{\langle[\hat{J}_i,\hat{J}_j]\rangle}{i\hbar}\,.
\end{equation}
If the operators are complete, any observable can be expressed as a function
of the expectation values $\langle\hat{J}_i\rangle$ and moments
\begin{equation}
 \Delta\left(\prod_iJ_i^{a_i}\right):= \left\langle\prod_i
 (\hat{J}_i-\langle\hat{J}_i\rangle)^{a_i}\right\rangle_{\rm symm}
\end{equation}
with operator products in totally symmetric ordering. Using linearity and the
Leibniz rule for Poisson brackets, these expectation values and moments form a
Poisson manifold. Their dynamics is determined by the Hamiltonian flow
generated by the expectation value $H_Q:=\langle\hat{H}\rangle$ of the
Hamiltonian constraint, another observable interpreted as a function of
expectation values and moments. Hamiltonian equations of motion usually couple
infinitely many moments to the expectation values, but a semiclassical
expansion to some finite order in $\hbar$ results in finitely coupled
equations which can be solved at least numerically. Computer-algebra codes
exist to automate the generation of equations to rather high orders
\cite{HigherMoments} (so far restricted to canonical commutators).

Writing $\hat{J}_i=\langle\hat{J}_i\rangle+
(\hat{J}_i-\langle\hat{J}_i\rangle)$ in the quantum Hamiltonian $H_Q=\langle
H(\hat{J}_i)\rangle$ and performing a formal expansion in
$(\hat{J}_i-\langle\hat{J}_i\rangle)$, the Hamiltonian flow is generated by
\begin{equation} \label{HQ}
 H_Q= H(\langle\hat{J}_i\rangle)+ \sum_{a_i} \frac{1}{a_1!}\cdots
 \frac{1}{a_N!} \frac{\partial^{a_1+\cdots a_N} 
   H(\langle\hat{J}_j\rangle)}{\partial
   \langle\hat{J}_1\rangle^{a_1}\cdots \partial\langle\hat{J}_N\rangle^{a_N}}
 \Delta\left(\prod_iJ_i^{a_i}\right)\,.
\end{equation}
The first term is the classical Hamiltonian evaluated in expectation values,
and the series includes quantum corrections of progressing order $\sum_i
a_i$. Equations of motion follow from Poisson brackets.

These constructions rely on commutators of linear operators and cannot be used
directly for non-linear Schr\"odinger-type equations. Nevertheless, a closely
related procedure can be followed for equations such as (\ref{NonLin}) in
which the non-linearity comes from non-local coefficients depending on the
moments. As one can readily confirm by computing time derivatives of
expectation values directly using (\ref{NonLin}) for wave-function factors,
the evolution of moments is now governed by a quantum Hamiltonian (\ref{HQ})
in which one initially treats the moments that appear in the non-local
coefficients as external functions; because they do not come from a linear
operator, they do not appear in commutators or in Poisson brackets of the
moments when equations of motion are derived. In the equations of motion, once
derived, these variables are to be equated to the moments they signify,
providing additional coupling terms between moments compared with a linear
Hamiltonian.

In our case, the $\hat{J}_i$ are given by three basic operators, a
multiplication operator by $n$ (the volume operator) and two shift operators
$\hat{h}$ and $\hat{h}^{\dagger}$ that change $n$ by $\pm 1$, implementing
(\ref{shift}) with $\delta=1$. In terms of canonical variables $(n,P)$, we can
write shift operators as quantizations of $h=\exp(iP)$. The commutators
\begin{equation}
 [\hat{n},\hat{h}]=-\hbar \hat{h} \quad,\quad [\hat{n},\hat{h}^{\dagger}]=
 \hbar\hat{h}^{\dagger} \quad,\quad [\hat{h},\hat{h}^{\dagger}]=0
\end{equation}
then define the basic algebra (\ref{JJ}) of our loop-quantized theory, and
correspondingly the Poisson brackets of expectation values and moments of $n$
and $h$. Moreover, since we introduced complex variables, the reality
condition $\hat{h}\hat{h}^{\dagger}=1$ as well as analogs for the moments
(such as $\Delta(hh^*)=1-hh^*$) must be satisfied.

We can realize the linear part of (\ref{NonLin}) as the Schr\"odinger equation
with Hamiltonian operator $\hat{h}+\hat{h}^\dagger-2$, resulting in the
quantum Hamiltonian $H_Q^{\rm lin}= h+h^*-2$ depending only on expectation
values but not on moments. Adding the non-linearity, we have an extra term
$-\frac{1}{2}\alpha A\langle\widehat{n^{-2}}\rangle$ with $A$ treated as a
constant to be set equal to $A=(\Delta n)^2$ in equations of motion, and
$\langle\widehat{n^{-2}}\rangle$ to be expanded by moments as in (\ref{HQ}).
For the difference equation (\ref{NonLin}), we then have the quantum
Hamiltonian
\begin{equation}
 H_Q=h+h^*-(2-\alpha A \langle \widehat{n^{-2}}\rangle)=
 h+h^*-2+\alpha A (3n^{-4}(\Delta n)^2- 20 n^{-5}\Delta(n^3)+\cdots)
\end{equation}
with the non-local coefficient $A$ treated for now as an external
parameter. (Instead of the inverse of $n$, which is ill-defined at $n=0$,
modifications due to inverse-triad corrections in loop quantum cosmology
should be used at small $n$ \cite{InvScale}.) 

We obtain the equations of motion from Poisson brackets, in which we then set
$A=(\Delta n)^2$:
\begin{eqnarray}
\dot{n} &=& i(h-h^*)\\
\dot{h} &=& 12i\alpha \frac{h}{n} \left(\frac{\Delta n}{n}\right)^4-6i\alpha
\frac{(\Delta   n)^2\Delta(nh)}{n^4}+\cdots\\
\frac{{\rm d}(\Delta n)^2}{{\rm d}t} &=& 2i(\Delta(nh)-\Delta(nh^*))
\end{eqnarray}
and so on for further moments.  Instead of $\ddot{n}=0$ as in the linear case,
we can combine the first two equations to obtain
\begin{equation}
 \ddot{n}= i(\dot{h}-\dot{h}^*)= -12\alpha\frac{h+h^*}{n} \left(\frac{\Delta
     n}{n}\right)^4- 6\alpha \frac{(\Delta n)^2(\Delta(nh)+\Delta(nh^*))}{n^4}
 +\cdots \,.
\end{equation}
Non-zero moments imply (negative) acceleration of the volume expansion.

\subsection{Interpretation}

Irrespective of the precise form of non-linearity, its presence has several
general consequences of potential importance for quantum cosmology. An obvious
and seemingly problematic implication is a loss of unitarity: wave functions
evolved by the non-linear equation do not have preserved scalar products with
other evolved states. There is no linear operator that could serve as a
Hamiltonian whose adjointness properties one could analyze by standard
techniques. Still, a straightforward direct calculation shows that the norms
$\langle\psi|\psi\rangle$ of states (but not scalar products
$\langle\phi|\psi\rangle$ of different states) are preserved. However, the
original many-body system is clearly unitary, and therefore non-unitarity is a
consequence of the reductions and approximations used. In order to interpret
the non-linearity correctly, we should therefore look back on the
constructions used to descend from many-body dynamics to a 1-particle
equation.

For a matter condensate, we obtain the non-linear wave equation (\ref{GP}) in
a rather indirect way: We do not reduce the many-body wave equation for $\Psi$
directly, but rather compute the expectation value of the Hamiltonian
(\ref{ExpH}), rewrite it in terms of the 1-particle wave function $\psi$, and
recognize the extra term as a formal analog of a potential depending on the
wave function. This potential, inserted in the standard Schr\"odinger equation,
then provides (\ref{GP}), a step which is again only formal. Experience shows
that the resulting non-linear equation nevertheless captures crucial
properties of the many-body problem, and rigorous proofs have been provided
\cite{KineticEqs,NonLinSchroed}.

One can avoid the last formal step by forgoing wave equations and instead
using the expectation value (\ref{ExpH}) to compute the spectrum of the
many-body Hamiltonian, for instance by variational methods applied to the
1-particle wave function $\psi$ on the right-hand side of (\ref{ExpH}). If the
spectrum of the Hamiltonian is known, evolution properties then follow without
directly using the non-linear equation (\ref{GP}). Similarly, effective
canonical equations in quantum mechanics refer to expectation values of the
Hamiltonian, such as (\ref{ExpH}) rather than wave equations, and are
therefore less sensitive to the apparent loss of unitarity. 

The physics of the system therefore does not suffer from a lack of
unitarity. Moreover, since the norm is still preserved, the probability
interpretation of a single state remains meaningful.  Instead of using
(\ref{GP}) as a fundamental wave equation for some function $\psi$ in a
Hilbert space, the equation models other dynamical effects, such as the
evolution of particle distributions or the approach and possible interaction
of superposed states. Properties such as the overlap of superposed states or
the distance between different distributions can be determined from moments of
a single wave function for the superposition and are independent of scalar
products of the wave function with other states; they can be analyzed with a
formal equation lacking unitarity. These are also the properties that
effective equations are sensitive to. In quantum cosmology, such questions are
usually of most interest because the exact state or wave function of quantum
space is not accessible by observations available now or in the foreseeable
future. Our model and with analog (\ref{NonLin}) of the Gross-Pitaevski
equation (\ref{GP}) is therefore reasonable.

\section{Discussion}

We have introduced a new model for inhomogeneous quantum cosmology, aiming to
capture essential features of the interacting dynamics of different parts of
quantum space. The processes we describe therefore provide the dynamics of
structure formation at a fundamental level. Using several approximations,
justified when inhomogeneity is sufficiently small, and importing ideas of
condensed-matter physics, we have been able to map the complicated many-body
dynamics to a non-linear minisuperspace equation.\footnote{Other versions of
  non-linear quantum cosmology have been proposed
  \cite{NonCommNonLin,QMNonLin,Parwani1,Parwani2}, motivated by
  non-commutativity and information-theoretic arguments.}

In addition to the approximate nature, several differences with the condensate
model occur:
\begin{itemize}
\item In the cosmological model, ``interactions'' between different patches
  are realized in superspace, not in actual space. Patches do not interact
  depending on their spatial distance, but depending on what their
  geometries are: The gravitational Hamiltonian depends on inhomogeneous
  modes, or on deviations of patch geometries from the spatial average. 
\item There is no delta-function potential (for pointlike interactions) but
  rather a polynomial potential, obtained by expanding the gravitational
  Hamiltonian as a function of patch geometries. As a consequence, the
  non-linearity is realized non-locally in the configuration space of wave
  functions.
\item While the many-body Hamiltonian of a condensate is well known but
  difficult to deal with, a consistent version of an inhomogeneous
  gravitational Hamiltonian in quantum gravity is still lacking. In
  particular, covariance conditions and the related problem of anomalies have
  not been evaluated in sufficient detail
  \cite{NonHerm,Komar,QSDI,LM:Vertsm,Consist}. (But see
  \cite{ThreeDeform,TwoPlusOneDef,TwoPlusOneDef2,AnoFreeWeak,AnoFreeWeakDiff}
  for recent progress.)
\end{itemize}

In this situation, having an approximate description of incompletely known
dynamics, we cannot expect to derive detailed quantitative cosmological
scenarios. (This statement does not only apply to our new method, but to all
derivations possible in quantum cosmology so far.) Effective techniques, as
used in our solution procedure for non-linear non-local equations, provide
means to parameterize ambiguities and ignorance, and to discuss anomalies, but
no details are available yet.  We therefore focus our discussion on new
qualitative features suggested by the non-linearity of the homogeneous model.

Non-linear wave equations provide new forms of minisuperspace effects that
capture crucial properties of averaged inhomogeneity. These terms need not
require high, near-Planckian densities to be significant because they could
potentially be large when many patch contributions are added up, even if each
of them is tiny. All leading contributions have the same sign because they
come from volume fluctuations, required to be positive. No cancellations
happen when one sums over all patches, potentially giving large effects.  For
certain behaviors of quantum fluctuations as functions of time or the volume,
our non-linearity can be interpreted as a cosmological-constant term, which
turns out to be negative. (Again, the sign is determined because quantum
fluctuations are always positive.) It remains to be seen whether more-refined
models, including those with anisotropic patches, or higher orders in the
moments in (\ref{WNonLin}), not all of which are restricted by positivity, as
well as perturbed Hamiltonians beyond second order can turn the sign to
provide an overall positive cosmological constant.

An interesting feature of non-linear wave equations is the existence of a
particular type of solutions: solitons. These are sharply peaked wave packets
which evolve without changing shape. Moreover, if solitons occur in
superposition, moving in different directions, they may occasionally overlap
but do not influence each other. After they have moved through the same spot,
they retain their old shapes. Such states are a promising candidate for new
dynamical coherent states in quantum cosmology. In contrast to kinematical
coherent states (or Gaussians) commonly used in such cases, solitons are
adapted to the dynamics and, in the indirect way that employs non-linear wave
equations, capture properties of inhomogeneity; in fact, their existence as
solutions relies on deviations from exact homogeneity.

The existence of solitons and the integrability of equations, together with
the associated possibility of chaos, depends sensitively on the form of
discrete equation \cite{NonLinChaos}. The discreteness, in turn, is related to
quantization and regularization ambiguities in canonical quantum gravity. The
strong sensitivity of some physical features may allow one to find tight
restrictions on ambiguities.

We end by mentioning another, more speculative consequence. In quantum
cosmology, solitons in superposition would correspond to different universes
superposed in the same state. Solitons may overlap but do not affect each
other's motion; they always form separate contributions to the total
state. Solitons and the non-linear wave equations they solve could therefore
play a role in the description and analysis of multiverse models.

\section*{Acknowledgements}

This work was supported in part by NSF grant PHY0748336.

%\bibliographystyle{preprint}
%\bibliography{Bib/QuantGra}

\end{document}